# The shear modulus of metastable amorphous solids with strong central and bond-bending interactions

*Alessio Zaccone*

Department of Chemistry and Applied Biosciences,

ETH Zurich, 8093 Zurich, Switzerland.

Email: alessio.zaccone@chem.ethz.ch.

Fax: 0041-44-6321082.

**Abstract**

We derive expressions for the shear modulus of deeply-quenched, glassy solids, in terms of a Cauchy-Born free energy expansion around a rigid (quenched) reference state, following the approach due to Alexander [Alexander, Phys. Rep. 296, 1998]. Continuum-limit explicit expressions of the shear modulus are derived starting from the microscopic Hamiltonians of central and bond-bending interactions. The applicability of the expressions to dense covalent glasses as well as colloidal glasses with strongly attractive or adhesive bonds is discussed.

# 1. Introduction

While the structure, elasticity, and lattice dynamics of condensed matter with long-range order (thanks to the intrinsic symmetry of crystalline structures) are fairly well understood [1], the same cannot be said of amorphous solids. Recent advances include the unveiling of connections between disordered solids made of thermal particles (glasses) and granular packings, so that the puzzling properties found in both these classes of materials can be investigated by means of unifying concepts. Well-known examples are the excess of low-frequency modes (the so-called Boson peak in the vibrational spectrum) [2], and the inhomogeneity of the elastic response [2-4]: features that have been observed in atomic (and molecular) glasses as well as in granular systems. These phenomena, as recent theoretical studies suggest, may find their origin in the weak connectivity of amorphous solids [5] as well as in their lack of symmetry [4,6]. Regarding the former aspect, recently it became clear that coordination plays a fundamental role in determining the mechanical properties of marginally-rigid solids when only central forces are at play. On the other hand, in the case of strongly connected structures or other dense systems where the bonds between building blocks can support bending moments, nonaffinity is often a very small correction to the affine part, thus the affine approximation works relatively well [7,8]. Some technologically important systems seem to belong to this class, e.g. dense networks of semi-flexible polymers, strong attractive colloidal glasses and covalent glasses (e.g., silicon glass) [7,9].

In the present work, we derive explicit expressions for the macroscopic shear modulus of deeply-quenched, arrested states of like particles, using Alexander's Cauchy-Born approach. The validity and application of the results are discussed.

## 2. Continuum theory of shear elasticity in solids with quenched disorder

In Ref. [6], S. Alexander formulated the systematic Cauchy-Born approach for amorphous solids, based on which the Helmholtz free energy at $T = 0$ (thus coinciding with the internal energy) can be expanded around a *rigid*, stressed, reference configuration where the set of particle positions is denoted by $\{R\}$. In such low-temperature reference state, as a result of quenching (solidification), particles are *labelled*, in the sense that they occupy well defined and fixed positions on a (disordered) lattice, the set of which represents just one out of $N!$ possible permutations ($N$ being the total number of particles). In other words, permutation symmetry (which is active in the liquid precursor) is broken in the quenching process [6]. Because of this, as opposed to equilibrium fluids, the disorder average for amorphous solids is of non-trivial definition. To avoid this problem, in Alexander's version of Cauchy-Born theory, the expansion (along with the disorder average) is carried out in terms of the *relative deviations* between particles. As shown in [6], this leads to the continuum limit and provides the only systematic application of Cauchy-Born theory to disordered solids. In the following we apply this approach to a generic dynamically arrested (glassy) state composed of spherical particles mutually interacting via two-body central and three-body angular (bond-bending) interactions. This may be a suitable model of well-bonded glassy systems such as atomic (covalent or metallic) glasses or attractive colloidal glasses.

Retaining terms up to second order, and including a three-body angular interaction term, the expansion reads

$$\delta F \equiv F(\{r\}) - F(\{R\}) \simeq \sum_{<ij>} \left.\frac{\partial F}{\partial r_{ij}}\right|_{\{R\}} \delta r_{ij} + \frac{1}{2}\sum_{<ij>} \left.\frac{\partial^2 F}{\partial r_{ij}^2}\right|_{\{R\}} (\delta r_{ij})^2 + \frac{1}{2}\sum_{<ijk>} \left.\frac{\partial^2 F}{\partial \Theta_{ijk}^2}\right|_{\{R\}} (\delta \Theta_{ijk})^2 \qquad (1.1)$$

In the first two terms on the r.h.s., the summation runs over all $N_c$ pairs of pair-interacting particles (i.e., over all bonds), and the derivatives are evaluated at the equilibrium distance in the reference rigid state $R_{ij} \equiv |\mathbf{R}_{ij}|$. In the last term (i.e. the angular interaction or bond-bending term) the summation is over pairs of bonds $[ij]$ and $[ik]$, $i \neq j \neq k$, having one common vertex. In our analysis, we will consider the two-body (central) terms and the three-body (bond-bending) terms in Eq. (1.1) separately, starting from the former case.

Expanding in the *relative* distance deviations allows one to define a microscopic displacement field $\mathbf{u}_{ij}$

$$\delta r_{ij} = u_{ij}^{\parallel} + [(u_{ij}^{\perp})^2 / 2R_{ij}] + \mathcal{O}(r_{ij}^3) \tag{1.2}$$

which has a component in the direction of $\mathbf{R}_{ij}$, i.e. $u_{ij}^{\parallel} \equiv (\delta \mathbf{R}_i - \delta \mathbf{R}_j) \cdot \hat{\mathbf{R}}_{ij}$, and an orthogonal component, i.e. $u_{ij}^{\perp} \equiv (\delta \mathbf{R}_i - \delta \mathbf{R}_j)^{\perp}$. In the absence of external forces, substituting Eq. (1.2) into the central-interaction terms in Eq. (1.1), gives

$$\delta F^{(C)} \simeq \sum_{<ij>} \left.\frac{\partial F}{\partial r_{ij}}\right|_{\{R\}} \frac{[(\delta \mathbf{R}_i - \delta \mathbf{R}_j)^{\perp}]^2}{2R_{ij}} + \frac{1}{2}\sum_{<ij>} \left.\frac{\partial^2 F}{\partial r_{ij}^2}\right|_{\{R\}} [(\delta \mathbf{R}_i - \delta \mathbf{R}_j) \cdot \hat{\mathbf{R}}_{ij}]^2 \tag{1.3}$$

The first-derivative terms correspond to the bond-tension or stress terms associated with the initial or quenched stresses (which are, generally, a non single-valued function of the aggregation history). These make an important contribution to the rigidity of weakly-connected materials and actually ensure the existence of a rigid reference state around which one can expand [6]. For central interactions, the second-derivative, can be written as the bond stiffness $\kappa_{\parallel} \equiv \left.\partial^2 F/\partial r_{ij}^2\right|_{\{R\}}$, also known as Born-Huang term. In the case of a glass, these terms have to be evaluated in the stressed state and in general may differ from the corresponding terms in a crystal at true thermodynamic equilibrium which are uniquely determined by the pair-interaction potential. The latter, however, could be still a reasonable approximation in the case of strong attractive glasses with a deep pair-interaction energy minimum [10]. Moreover, it has been shown that in glasses the stress terms in the expansion, Eq. (1.3), do not contribute to the macroscopic elastic moduli which are, therefore, solely determined by the Born-Huang terms [6,11]. Thus, one can write

$$\delta F^{(C)} \simeq \frac{1}{2}\kappa_{\parallel}\sum_{<ij>}\overline{[(\delta \mathbf{R}_i - \delta \mathbf{R}_j)\cdot \hat{\mathbf{R}}_{ij}]^2} = \frac{1}{2}\kappa_{\parallel}\sum_{<ij>}\overline{(u_{ij}^{\parallel})^2} \qquad (1.4)$$

where $\overline{\bullet}$ denotes the average over all possible *deviations* (i.e. strain configurations) from the reference state. Introducing a smooth continuum displacement field $\mathbf{u}(\mathbf{r})$, to lowest order in thr gradient expansion one has

$$\overline{u_{ij}^{\parallel}} \simeq (\boldsymbol{R}_{ij}\cdot \nabla)\mathbf{u}(\mathbf{r})\cdot \hat{\boldsymbol{R}}_{ij} = R_{ij}^{-1}R_{ij}^{\alpha}R_{ij}^{\beta}\partial_{\alpha}u_{\beta}, \qquad (1.5)$$

where summation over repeated indices is understood and transposition symmetry is evident. Using this and introducing the affine transformation $\overline{(u_{ij}^{\parallel})^2} \simeq \text{Tr}[(\boldsymbol{R}_{ij}\otimes \boldsymbol{R}_{ij})\cdot \mathbf{e}/R_{ij}]^2$ defined by the disorder-averaged linearized symmetric strain tensor $\mathbf{e} \equiv e_{\alpha\beta} = \tfrac{1}{2}(\partial_{\alpha}u_{\beta}+\partial_{\beta}u_{\alpha})$, with $\overline{\overline{u_{ij}^{\parallel}}^2 - \overline{u_{ij}^{\parallel}}^2} \ll \overline{u_{ij}^{\parallel}}^2$, we obtain the continuum limit [6]:

$$\delta F^{(C)} \simeq \frac{1}{2}\sum_{<ij>}\kappa_{\parallel}\overline{(u_{ij}^{\parallel})^2} \simeq \frac{1}{2}\sum_{<ij>}\kappa_{\parallel}\left\{\frac{\text{Tr}[(\boldsymbol{R}_{ij}\otimes \boldsymbol{R}_{ij})\cdot \mathbf{e}]}{R_{ij}}\right\}^2 \qquad (1.7)$$

where $\otimes$ denotes the dyadic product. It is easy to find that for an imposed pure shear deformation the above expression reduces to

$$\delta F^{(C)} \simeq \frac{1}{2}\kappa_{\parallel}\sum_{<ij>}4R_{ij}^2\left(\frac{R_{ij}^x}{R_{ij}}\frac{R_{ij}^y}{R_{ij}}\right)^2 e_{xy}^2 \qquad (1.8)$$

For the quenched configuration $\{R\}$, under the assumption that pair (two-body) interactions are much stronger than higher-order multi-body interactions, the summation over pairs of nearest-neighbours can be replaced by the total number of bonds, $N_c$. This implies an average over all possible spatial orientations of the bonds in the reference state $\{R\}$:

$$\delta F^{(C)} \simeq \frac{1}{2} \kappa_{||} N_c \left\langle 4 R_{ij}^2 (\hat{R}_{ij}^x \hat{R}_{ij}^y)^2 e_{xy}^2 \right\rangle_\Omega \simeq 2 \kappa_{||} N_c R_0^2 \left\langle (\hat{R}_{ij}^x \hat{R}_{ij}^y)^2 \right\rangle_\Omega e_{xy}^2 \tag{1.9}$$

where $R_{ij} \equiv R_0$ is the average interparticle distance in the reference quenched configuration and $\langle \bullet \rangle_\Omega$ denotes the angular average. Introducing the mean coordination $z$, and noting that $N_c/V \equiv \frac{1}{2} z N/V \equiv 3 z \phi / \pi R_0^d$, leads to the following form for the free energy density

$$\delta \tilde{F}^{(C)} \simeq 6 \pi^{-1} \kappa_{||} z \phi R_0^{2-d} \left\langle (\hat{R}_{ij}^x \hat{R}_{ij}^y)^2 \right\rangle_\Omega e_{xy}^2. \tag{1.10}$$

where $d$ is the dimensionality of space. Hence, using spherical coordinates $\hat{\mathbf{R}}_{ij} = (\sin\theta \cos\varphi, \sin\theta \sin\varphi, 1)$, with $R_{ij}^x = \sin\theta \cos\varphi$, $R_{ij}^y = \sin\theta \sin\varphi$. Assuming that the particles have zero degree of spatial correlation, averaging gives

$$\left\langle (\hat{R}_{ij}^x \hat{R}_{ij}^y)^2 \right\rangle_\Omega = \frac{1}{4\pi} \int \int d\varphi \sin\theta d\theta (\sin^4\theta \cos^2\varphi \sin^2\varphi) = \frac{1}{15} \tag{1.11}$$

At $T=0$, $\sigma_{\alpha\beta} \equiv \partial \delta F / \partial e_{\alpha\beta}$ and the affine translation-rotation invariant shear modulus for the central-force case can be derived as

$$G^{(C)} \simeq \frac{4}{5\pi} \kappa_{||} z \phi R_0^{2-d}. \tag{1.12}$$

The coordination number $z$ can be estimated from the experimentally determined structure factor, or evaluated, for sufficiently dense glasses, according to the following route. If the glass is dense ($\phi > 0.5$) its structure is homogenous due to mutual impenetrability of the particles and therefore dominated by the hard-sphere component of interaction. As shown by recent experimental studies [12], the result is that dense ($\phi \sim 0.6$) strongly attractive glasses exhibit the same homogeneous structure of purely hard-sphere glasses. Therefore, it is possible to estimate the mean coordination as a function of the packing fraction $\phi$, by calculating the mean coordination of the *hyper-quenched* hard-sphere liquid with the same $\phi$. This is equivalent to integrating the radial distribution function of hard-sphere liquids with a cut-off on the integration determined so as to recover the jamming point of monodisperse hard-spheres (given as $z=6$ at $\phi \simeq 0.64$). This route has been used to interpret experimental data of attractive colloidal glasses in [11].

Eq. (1.12) has been obtained under the limiting assumption that, in very attractive systems, the affine approximation leads to a small error. However, in spite of that approximation, in [11] it has been shown that Eq. (1.12) gives a rather accurate, quantitative description of the shear modulus of short-ranged attractive (depletion) colloidal glasses such as those studied in [13]. In that case, the affine approximation is justified because the elastic response is dominated by the first linear regime ending with break-up of nearest-neighbour bonds [13].

The more general expansion in Eq. (1.1) involves the three-body bond-bending forces and is somewhat more complex. A suitable model, which satisfies translation-rotation invariance, is the three-body Hamiltonian [14]

$$\delta F^{(C)} = \frac{1}{2}\kappa_\perp \sum_{<ijk>} (\delta\Theta_{ijk})^2 = \frac{1}{2}\kappa_\perp \sum_{<ijk>} \left|(\mathbf{u}_{ij}\times\hat{\mathbf{R}}_{ij} - \mathbf{u}_{ik}\times\hat{\mathbf{R}}_{ik})\cdot(\hat{\mathbf{R}}_{ij}\times\hat{\mathbf{R}}_{ik})/\left\|\hat{\mathbf{R}}_{ij}\times\hat{\mathbf{R}}_{ik}\right\|\right|^2 \qquad (1.13)$$

where $\kappa_\perp$ is the local BB stiffness: $\kappa_\perp \equiv \partial^2 F/\partial\Theta_{ijk}^2\big|_{\{R\}}$. Again, following the Cauchy-Born approach of [6] and averaging over all possible strained configurations one can write

$$\left|\hat{\mathbf{R}}_{ij}\times\hat{\mathbf{R}}_{ik}\right|\overline{\delta\Theta_{ijk}} = \left(\overline{\mathbf{u}_{ij}}\times\hat{\mathbf{R}}_{ij} - \overline{\mathbf{u}_{ik}}\times\hat{\mathbf{R}}_{ik}\right)\cdot\left(\hat{\mathbf{R}}_{ij}\times\hat{\mathbf{R}}_{ik}\right), \qquad (1.14)$$

which, in component notation, and after expanding in the displacement field, reads

$$\left|\hat{\mathbf{R}}_{ij}\times\hat{\mathbf{R}}_{ik}\right|\overline{\delta\Theta_{ijk}} \simeq (R_{ij}^{-1}R_{ij}^{\alpha}\partial_{\alpha}\varepsilon_{\beta\delta\gamma}u_{\delta}R_{ij}^{\gamma} - R_{ik}^{-1}R_{ik}^{\chi}\partial_{\chi}\varepsilon_{\beta\eta\lambda}u_{\eta}R_{ik}^{\lambda})\varepsilon_{\beta\mu\nu}R_{ij}^{\mu}R_{ij}^{\nu} \tag{1.15}$$

As shown in the Appendix, one has that

$$\left(\overline{\mathbf{u}_{ij}}\times\hat{\mathbf{R}}_{ij}\right)\cdot\left(\hat{\mathbf{R}}_{ij}\times\hat{\mathbf{R}}_{ik}\right) \simeq 2\left[(\mathbf{R}_{ij}^{\mathrm{T}}\cdot\mathbf{e})\times\hat{\mathbf{R}}_{ij}\right]\cdot\left(\hat{\mathbf{R}}_{ij}\times\hat{\mathbf{R}}_{ik}\right). \tag{1.16}$$

Thus, the disorder-averaged change in the interaction angle can be written as

$$\overline{\delta\Theta_{ijk}} \simeq 2\left[(\mathbf{R}_{ij}^{\mathrm{T}}\cdot\mathbf{e})\times\hat{\mathbf{R}}_{ij} - (\mathbf{R}_{ik}^{\mathrm{T}}\cdot\mathbf{e})\times\hat{\mathbf{R}}_{ik}\right]\cdot(\hat{\mathbf{R}}_{ij}\times\hat{\mathbf{R}}_{ik})/\left|(\hat{\mathbf{R}}_{ij}\times\hat{\mathbf{R}}_{ik})\right| \tag{1.17}$$

which, by making use of Lagrange's identity and rearranging terms, becomes

$$\overline{\delta\Theta_{ijk}} \simeq 2(\sin\Theta_{ijk})^{-1}\left\{\left[(\mathbf{R}_{ij}^{\mathrm{T}}\cdot\mathbf{e})\cdot\hat{\mathbf{R}}_{ij} + (\mathbf{R}_{ik}^{\mathrm{T}}\cdot\mathbf{e})\cdot\hat{\mathbf{R}}_{ik}\right]\cos\Theta_{ijk} - \left[(\mathbf{R}_{ij}^{\mathrm{T}}\cdot\mathbf{e})\cdot\hat{\mathbf{R}}_{ik} + (\mathbf{R}_{ik}^{\mathrm{T}}\cdot\mathbf{e})\cdot\hat{\mathbf{R}}_{ij}\right]\right\} \tag{1.18}$$

and, finally,

$$\overline{\delta\Theta_{ijk}} \simeq 2(R_0\sin\Theta_{ijk})^{-1}\left\{\cos\Theta_{ijk}\left[\mathrm{Tr}(\mathbf{R}_{ij}\otimes\mathbf{R}_{ij}) + \mathrm{Tr}(\mathbf{R}_{ik}\otimes\mathbf{R}_{ik})\right]\cdot\mathbf{e} - \left[\mathrm{Tr}(\mathbf{R}_{ik}\otimes\mathbf{R}_{ij}) + \mathrm{Tr}(\mathbf{R}_{ij}\otimes\mathbf{R}_{ik})\right]\cdot\mathbf{e}\right\}$$

$$\tag{1.19}$$

For a pure shear, the above expression reduces to

$$\overline{\delta\Theta_{ijk}} \simeq 4(R_0\sin\Theta_{ijk})^{-1}\left\{\left[(R_{ij}^{x}R_{ij}^{y} + R_{ik}^{x}R_{ik}^{y})\cos\Theta_{ijk}\right] - \left[R_{ij}^{y}R_{ik}^{x} + R_{ij}^{x}R_{ik}^{y}\right]\right\}e_{xy} \tag{1.20}$$

We now take the isotropic average over $\Theta_{ijk}$, thus assuming a flat distribution for $\Theta_{ijk}$: this assumption may be realistic for systems with strong spatial disorder such as e.g. emulsion glasses, colloidal or atomic (metallic and semiconductor) glasses without directional interactions. For molecular network glasses, however, $\Theta_{ijk}$ will rather be distributed according to the chemistry of the system. With covalent network-glasses, usually the number of angles $\Theta_{ijk}$ is finite and dictated by the valence, thus giving rise to distinct terms in the expansion. Application of this model to specific covalent glasses may be the object of future work. Here we limit our analysis to the case of strong disorder, so that an unbiased average yields

$$\left\langle \overline{\delta\Theta_{ijk}} \right\rangle_\Theta \simeq \left\langle 4(R_0 \sin\Theta_{ijk})^{-1} \left\{ \left[ (R_{ij}^x R_{ij}^y + R_{ik}^x R_{ik}^y) \cos\Theta_{ijk} \right] - \left[ R_{ij}^y R_{ik}^x + R_{ij}^x R_{ik}^y \right] \right\} \right\rangle_\Theta e_{xy}$$
$$\simeq \frac{4}{3} R_0 \sin\varphi \cos\varphi \, (-\sin^2\theta + \cos^2\theta - 4\sin\theta\cos\theta) e_{xy} \quad (1.21)$$

where $R_{ik}^x = \sin(\theta+\Theta)\cos\varphi$ and $R_{ik}^y = \sin(\theta+\Theta)\sin\varphi$ have been used. As before, we assume random orientation of the bonds in the reference state $\{R\}$ so that on average each term in the summation in Eq. (1.13) contributes

$$\left\langle \left\langle \overline{\delta\Theta_{ijk}} \right\rangle_\Theta^2 \right\rangle_\Omega \simeq \frac{16}{9} R_0^2 \left[ \frac{1}{4\pi} \int\int d\varphi \sin\theta d\theta (\sin\varphi\cos\varphi)^2 \right.$$
$$\left. \times (\sin^4\theta + \cos^4\theta + 14\sin^2\theta\cos^2\theta + 8\sin^3\theta\cos\theta - 8\sin\theta\cos^3\theta) \right] e_{xy}$$

(1.22)

Therefore, using $\overline{\delta\Theta_{ijk} - \overline{\delta\Theta}_{jk}}^2 \ll \overline{\delta\Theta_{ijk}}^2$ for the average over disorder, as well as $\left\langle \overline{\delta\Theta_{ijk}} - \left\langle \overline{\delta\Theta_{ijk}} \right\rangle_\Theta \right\rangle_\Theta^2 \ll \left\langle \overline{\delta\Theta_{ijk}} \right\rangle_\Theta^2$ for the spatial average over the bending angle, linear elasticity leads to

$$G^{(B)} \simeq \frac{124}{135\pi} \kappa_\perp z^{(B)} \phi R_0^{2-d} \quad (1.23)$$

where the sum over three-body interactions has been replaced by $\frac{1}{3} zN/V \equiv 2z\phi/\pi R_0^d$.

In Eq. (1.12) and (1.23) the definition of the microscopic bond rigidities ($\kappa_\parallel$ and $\kappa_\perp$, respectively) is clearly different, and the numerical prefactor is also different. In the BB case, the value of the prefactor is especially important, because for (real) network-glasses it also contains information about the chemistry-dependent geometry of the network. Here, the prefactor $(124/135)\pi^{-1}$ has been found for the case of nondirectional bonds and strong disorder but, in the case of real covalent glasses, it will depend on the values of the bond-bending angle $\Theta_{ijk}$. For a generic system where both CF and BB interactions are present, as in a real glass, the shear modulus can be estimated as

$$G = G^{(C)} + G^{(B)} \simeq \left(\frac{4}{5\pi}\kappa_{\parallel}z^{(C)} + \frac{124}{135\pi}\kappa_{\perp}z^{(B)}\right)\phi R_0^{2-d} \qquad (1.24)$$

Eq. (1.24) accounts for the fact that the mean number of bonds per particle which display BB resistance may differ from that of purely CF bonds. Indeed, for real covalent glasses, $z^{(B)}$ is a function of the valence which, in turn, is determined by the specific chemistry of the glass under consideration.

## 3. Discussion and potential applications

Eqs. (1.12) and (1.23) have been derived by systematically applying Cauchy-Born theory (with the expansion written in terms of the *relative* deviations) and give the macroscopic elastic response to shear of amorphous solids with central-force and bond-bending interactions, respectively, as a function of coarse-grained parameters. These are the mean coordination ($z$), the volume fraction ($\phi$), the interparticle interactions (embedded in the Born-Huang term $\kappa$), and the mean separation distance ($R_0$) between nearest-neighbours in the reference (stressed) configuration. The latter, in a solid, is approximately equal to the diameter of the building blocks. We would like to clarify, at this point, the differences of the approach outlined in this work as compared to the various models in the literature, especially within rheological models. In most of these cases, the heterogeneous and heuristic character of the assumptions leads to numerical prefactors inconsistent with each other and generally not comparable with the experiments [15]. Moreover, the affine approximation is sometimes applied to weakly bonded materials where nonaffine rearrangements are instead important. Including bond-bending terms in the expansion as done here is crucial thus making possible the application of continuum theory to such materials as strong covalent glasses where nonaffine rearrangements are smaller [9].

Finally, we note that Eq. (1.23) and (1.24) may also find application in understanding the structure-elasticity properties of dense aggregated colloidal systems. In fact, it has been recently shown that polymer latex particles in the micron range display BB rigidity as a consequence of contact adhesion [16,17]. Therefore, for such colloidal systems, the BB stiffness in Eq. (1.22) may be expressed as a function of the surface adhesion parameters according to the experimental findings of [16], where the relation $\kappa_\perp \simeq 6\pi a_c^4 E_0/R_0^3$ was proposed ($a_c$ is the radius of the contact area of adhesion between two particles and $E_0$ is the particle Young's modulus). Our microscopic definition of $\kappa_\perp$ is fully consistent with the one given in Refs. [16] and [18], thus making our model potentially applicable to aggregated colloidal systems. Also in the case where the colloidal bonds are not adhesive, provided they are sufficiently short-ranged and of considerable attractive strength, predictions based on Eq. (1.12) are in quantitative agreement with experimentally measured values, as shown in [11].

## 4. Conclusions

The systematic Cauchy-Born approach to amorphous solids, in the same spirit of Ref. [6], has been applied to evaluate the macroscopic response to shear of low-$T$ glassy states of spherical particles interacting via a central pair-interaction potential supplemented with an angular (bond-bending) three-body interaction term. Expressions in closed form are derived by making use of the affine approximation. The latter is generally a strong assumption when dealing with disordered systems, but may lead to small errors if the interparticle bonds can support significant bending moments (thus greatly reducing the number of degrees of freedom), as in covalent glasses (e.g. chalcogenide and amorphous silicon) [9]. Further, the model has the potentiality to account for the specific chemistry-dependent structure of real glasses. In the case of purely *central* pair interaction potentials, the situation is more complex because nonaffine relaxations are usually important. The affine approximation, therefore, is of limited application. However, also in the latter case, as shown elsewhere [11], the formulae derived here can nevertheless yield accurate predictions for colloidal glasses in the limit of strong short-ranged interparticle attraction. In this limit, the observed linear elastic regime is indeed due to stretching of the bonds [13,19], so that, the particles being localized upon strain within the short range of attraction, the assumptions used here yield reasonable predictions.

**Acknowledgments**

Financial support of the Swiss National Science Foundation (grant. No. 200020-113805/1) is gratefully acknowledged. I am grateful to Dr. Hua Wu and Prof. Massimo Morbidelli for support and discussions. I am indebted to Dr. Emanuela Del Gado for her support and critical reading of the manuscript, and to Dr. Marco Lattuada for discussions.

**Appendix. Derivation of Eq. (1.16)**

One can decompose the gradient expansion of the smooth displacement field $\mathbf{u}(\mathbf{r})$ into an explicitly symmetric part (i.e. the disorder-averaged symmetric strain tensor) and an antisymmetric one as:

$$\overline{\mathbf{u}_{ij}} \equiv \overline{\mathbf{u}_i - \mathbf{u}_j} \simeq (\mathbf{R}_{ij} \cdot \nabla)\mathbf{u} = \left[ \mathbf{R}_{ij}^{\mathrm{T}} \cdot \mathbf{e} + \frac{1}{2}(\nabla \times \mathbf{u}) \times \mathbf{R}_{ij} \right] \tag{A.1}$$

Using the well-known identities:

$$(\nabla \times \mathbf{u}) \times \mathbf{R}_{ij} = -\mathbf{R}_{ij} \times (\nabla \times \mathbf{u}) = -\nabla(\mathbf{R}_{ij} \cdot \mathbf{u}) + (\mathbf{R}_{ij} \cdot \nabla)\mathbf{u} \tag{A.2}$$

Eq. (A.1) can be rewritten as:

$$\overline{\mathbf{u}_{ij}} \simeq (\mathbf{R}_{ij} \cdot \nabla)\mathbf{u} = \left[ \mathbf{R}_{ij}^{T} \cdot \mathbf{e} - \frac{1}{2}\nabla(\mathbf{R}_{ij} \cdot \mathbf{u}) + \frac{1}{2}(\mathbf{R}_{ij} \cdot \nabla)\mathbf{u} \right] \tag{A.3}$$

Rearranging terms:

$$\overline{\mathbf{u}_{ij}} \simeq (\mathbf{R}_{ij} \cdot \nabla)\mathbf{u} = 2\left[ \mathbf{R}_{ij}^{T} \cdot \mathbf{e} - \frac{1}{2}\nabla(\mathbf{R}_{ij} \cdot \mathbf{u}) \right] \tag{A.4}$$

Hence, we can rewrite Eq. (16) as:

$$\left(\overline{\mathbf{u}_{ij}} \times \hat{\mathbf{R}}_{ij}\right) \cdot \left(\hat{\mathbf{R}}_{ij} \times \hat{\mathbf{R}}_{ik}\right) \simeq 2\left\{[(\mathbf{R}_{ij}^{T} \cdot \mathbf{e}) \times \hat{\mathbf{R}}_{ij}] - \frac{1}{2}\left[\nabla\left(\mathbf{R}_{ij} \cdot \mathbf{u}\right) \times \hat{\mathbf{R}}_{ij}\right]\right\} \cdot \left(\hat{\mathbf{R}}_{ij} \times \hat{\mathbf{R}}_{ik}\right) \tag{A.5}$$

For the antisymmetric parts, we can use the identity $\nabla(\mathbf{R}_{ij} \cdot \mathbf{u}) \times \hat{\mathbf{R}}_{ij} = \nabla \times (\mathbf{R}_{ij} \cdot \mathbf{u})\hat{\mathbf{R}}_{ij} - (\mathbf{R}_{ij} \cdot \mathbf{u})\nabla \times \hat{\mathbf{R}}_{ij}$, where the second term on the RHS is clearly zero. Therefore, the term $[\nabla \times (\mathbf{R}_{ij} \cdot \mathbf{u})\hat{\mathbf{R}}_{ij}] \cdot (\hat{\mathbf{R}}_{ij} \times \hat{\mathbf{R}}_{ik})$, making use of Lagrange's identity, is seen to be zero

$$[\nabla \times (\mathbf{R}_{ij} \cdot \mathbf{u})\hat{\mathbf{R}}_{ij}] \cdot (\hat{\mathbf{R}}_{ij} \times \hat{\mathbf{R}}_{ik}) = (\nabla \cdot \hat{\mathbf{R}}_{ij})[(\mathbf{R}_{ij} \cdot \mathbf{u})\hat{\mathbf{R}}_{ij} \cdot \hat{\mathbf{R}}_{ik}] - (\nabla \cdot \hat{\mathbf{R}}_{ik})[(\mathbf{R}_{ij} \cdot \mathbf{u})\hat{\mathbf{R}}_{ij} \cdot \hat{\mathbf{R}}_{ij}] = 0 \tag{A.6}$$

because $\hat{\mathbf{R}}_{ij}$ and $\hat{\mathbf{R}}_{ik}$ are constant vectors. Hence, Eq. (1.16) is verified.

**References**

[1] Born M and Huang H 1954 *Dynamical Theory of Crystal Lattices* (Oxford: Oxford University Press).

[2] O'Hern C S, Silbert L E, Liu A J, and Nagel S R 2003 Phys. Rev.E **68** 011306.

[3] DiDonna B A, and Lubensky T C, Physical Review E 2005 **72** 23.

[4] Lemaitre A, and Maloney C, Journal of Statistical Physics 2006 **123**, 415; Maloney C E, Physical Review Letters 2006 **97** 4.

[5] Wyart M, Annales de Physique 2005 **30** 1; arXiv:0806.4653v1.

[6] Alexander S, Physics Reports 1998 **296** 65.

[7] Das M, MacKintosh F C, and Levine A J, Physical Review Letters 2007 **99** 4.

[8] Zaccone A *et al.*, Journal of Chemical Physics 2007 **127** 174512.

[9] Wyart *et al.*, Physical Review Letters 2008 **101** 215501.

[10] Tanguy A *et al.*, Physical Review B 2002 **66**, 17.

[11] Zaccone A, Wu H, and Del Gado E, arXiv:0901.4713v1.

[12] Kaufman L J and Weitz D A, Journal of Chemical Physics 2006 **125** 074716.

[13] Pham K N *et al.*, Journal of Rheology 2008 **52** 649; K. N. Pham *et al.*, Europhysics Letters 2006 **75** 624.

[14] Sahimi M, Physics Reports-Review Section of Physics Letters 1998 **306** 214.

[15] Buscall R, et al., Journal of the Chemical Society-Faraday Transactions I 1982 **78** 2889; Buscall R, Journal of the Chemical Society-Faraday Transactions 1991 **87** 1365.

[16] Pantina J P and Furst E M, Physical Review Letters 2005 **94** 138301; Pantina J P and Furst E M, Langmuir 2006 **22** (12) 5282.

[17] Becker V and Briesen H, Physical Review E 2009 **71** 061404;

[18] Kantor Y, and Webman I, Physical Review Letters 1998 **52** 1891.

[19] Pusey P N J. Phys.: Condens. Matter 2008 20 494202.